\documentclass{article}

\PassOptionsToPackage{numbers, compress}{natbib}

\usepackage[preprint]{neurips_2026}
\usepackage{dashrule}
\usepackage[utf8]{inputenc}   
\usepackage[T1]{fontenc}      
\usepackage{url}              
\usepackage{microtype}        
\usepackage{nicefrac}         

\usepackage{amsmath, amsfonts, amssymb, amsthm, mathtools}

\usepackage{booktabs}         
\usepackage{multirow}         
\usepackage{makecell}         
\usepackage{array}            
\usepackage{adjustbox}        
\usepackage[table]{xcolor}    

\usepackage{graphicx}
\usepackage{subcaption}
\usepackage{rotating}         

\usepackage{hyperref}
\usepackage{cleveref}         

\usepackage[disable]{todonotes} 
\usepackage{enumitem}         
\usepackage[skins, breakable]{tcolorbox} 

\definecolor{titleblue}{RGB}{240,247,255}
\definecolor{warnred}{RGB}{180,0,0}

\title{StrixAE: An Intelligent Agent for Audio Enhancement under Complex Distortion Coupling in Real-World Scenarios}

\author{%
Chenglin Wu$^{1,*}$ \quad
Junjie Wu$^{1,*}$ \quad
Jinhong Chen$^{1}$ \quad
Mingyang Chen$^{1}$ \\
Zixu Lin$^{1}$ \quad
Jiabian Chen$^{2}$ \quad
Xinghao Ding$^{1}$ \quad
Xiaotong Tu$^{1}$ \\[4pt]
$^{1}$Xiamen University \\
$^{2}$Fuyao University of Science and Technology \\
$^{*}$Equal contribution.
}

\begin{document}
\maketitle
\begin{abstract}
Audio enhancement in real-world scenarios involves complex distortion couplings and requires personalized enhancement. Existing solutions struggle to address both simultaneously. To improve robustness andenable autonomous operation in such scenarios, we propose StrixAE, an agent based on a multimodal large language model (MLLM). StrixAE leverages the MLLM as a controller to coordinate multiple audio enhancement and personaliza-tion models. To further enhance system robustness, reduce artifacts, and improve generalization across diverse real-world scenarios, StrixAE is trained through a two-stage process: first, CoT supervised fine-tuning on AcoustBench to ground basic reasoning and tool invocation; second, Audio Perception Reinforcement Learning (APRL) — a reward design specifically tailored for audio restoration pipelines that jointly optimizes format validity, structural coherence, and perceptual quality. Unlike generic RL fine-tuning, APRL introduces structured rewards that enforce executable pipelines and logical section ordering, enabling the agent to produce reliable, interpretable enhancement plans without hallucinated tools.
Based on real-world test datasets, our proposed method outperforms most existing open-source and proprietary solutions, achieving state-of-the-art performance across multiple perception metrics and demonstrating strong generalization robustness.
\end{abstract}

\section{Introduction}

Speech enhancement (SE) aims to improve speech quality and intelligibility while preserving the intrinsic characteristics of the original signal \cite{maher2018principles}. In real-time communication, speech quality is significantly compromised by background noise \cite{reddy2020interspeech}, reverberation \cite{kinoshita2016summary}, and interfering speakers \cite{saijo2025interspeech,li2026icassp}. This field encompasses tasks such as noise reduction \cite{wang2025mamba}, dereverberation \cite{lu2023mp}, speech separation \cite{zhao2025mossformer2,xu2024tiger}, and bandwidth extension \cite{ravanelli2024open}. However, current audio enhancement models are predominantly categorized into task-specific methods \cite{wang2025mamba,lu2023mp,chao2024investigation,cao2022cmgan} and ensemble methods \cite{zhao2025clearervoice,dubey2024icassp,ju2023tea,lu2023mp,ju2022tea,wang2023tf}, often trained on limited datasets \cite{lu2023mp,wang2025mamba,dubey2024icassp}, lacking a standardized benchmark for versatility and a holistic architecture to handle diverse acoustic conditions.

\begin{figure}[t]
  \begin{center}
    \centerline{\includegraphics[width=0.88\columnwidth]{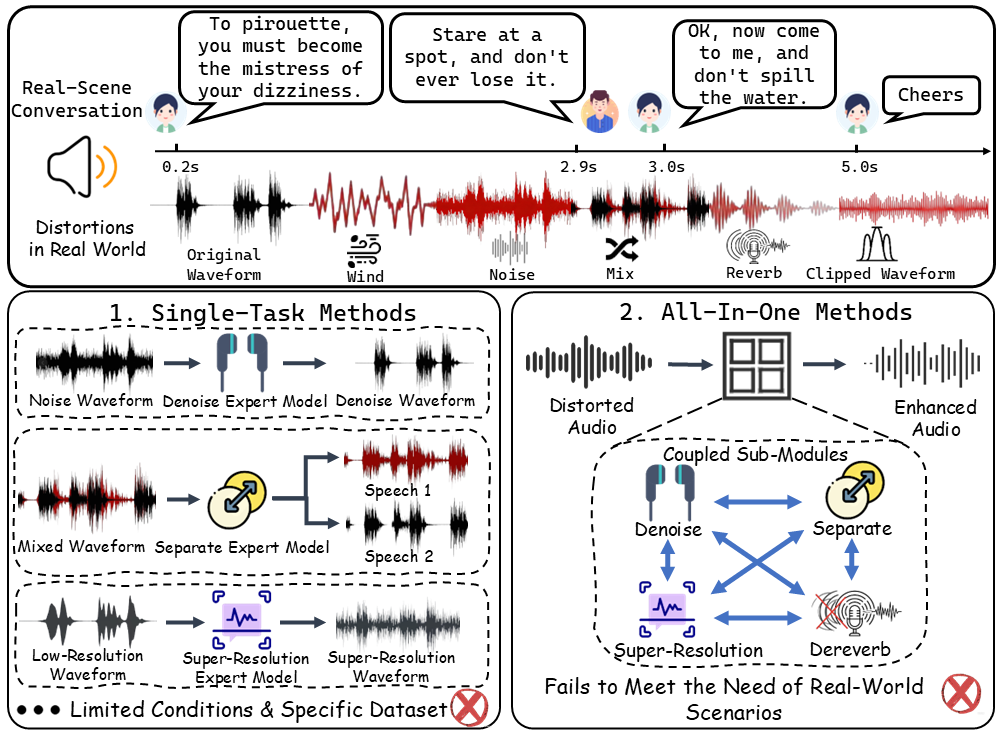}}
    \caption{
        Limitations of Single-Task and All-In-One methods: (1) Single methods only focus on limited distortion conditions and augmentation on specific datasets. (2) All-In-One methods cannot address the needs of coupled distortion and personalized augmentation in real-world scenarios. }
    \label{fig1}
  \end{center}
\end{figure}

Conventional task-specific methods \cite{lu2023mp,wang2025mamba,dubey2024icassp} primarily focus on single-objective speech enhancement, such as denoising or dereverberation. Although ensemble-based strategies extend this scope to composite degradations, they often fail in complex environments that require selective enhancement or suppression of interfering speakers. The inherent unpredictability and coupled complexity of real-world audio further challenges the generalization ability of existing models, as illustrated in Figure~\ref{fig1}. Consequently, there remains a critical need for standardized benchmarks that can evaluate system versatility, as well as holistic architectures capable of handling diverse acoustic conditions across multiple scenarios.

Recently,  Multimodal large language model (MLLM) have achieved breakthroughs in reasoning and autonomous task execution \cite{jain2022introduction,yang2023gpt4tools,zhao2024see,lin2025jarvisir,feng2026gen}. Because the type and combination of distortions are unknown a priori, any fixed pipeline or static model fails. This naturally calls for a reasoning-capable controller, which can be provided by  MLLM. Inspired by this, we explore a novel audio enhancement paradigm featuring proactive perception of real-world distortions, interpretable handling of composite degradation, and professional-grade enhancement. While MLLMs demonstrate inherent audio perception capabilities as controllers for expert models, constructing such an agent requires extensive paired supervisory data. The scarcity of labeled distorted audio hinders supervised fine-tuning \cite{zhang2024neurips,saijo2025interspeech}. To address these issues, we construct a multi-scenario dataset with coupled degradations and curate 88.9K supervised samples.

In this paper, we propose StrixAE, an audio agent based on multimodal large-scale models (MLLMs), leveraging Audio-Reasoner  \cite{xie2025audio} and expert models from open-source communities. The system development comprises two key components: (1) AcoustBench, an instruction-following dataset built via a self-guided strategy \cite{wang2023self}, containing \textbf{200} hours of personalized DNS (PDNS) and dual/single-speaker audio, \textbf{88.9K} synthetic instruction-response pairs, and \textbf{1.9K} real-world test pairs; (2) A training framework combining Supervised Fine-Tuning (SFT) and AI feedback alignment to enhance reliability and autonomy. We fine-tune StrixAE using AcoustBench to boost robustness, reduce hallucinations, and improve generalization and instruction alignment. To ensure stable training, we propose GRPO-AE, incorporating ranking-based supervision inspired by Ranked Responses from Human Feedback (RRHF). In addition, we integrate multiple audio quality assessment models into a unified reward model to provide comprehensive feedback.

The main contributions of our work can be summarized as follows:
\begin{itemize}

\item We propose StrixAE, a novel MLLM-based audio agent capable of perceiving degradationfactors and autonomously orchestrating expert models for multi-scenario audio enhancement.

\item \textbf{Two-stage training paradigm for audio agents.} We propose SFT followed by APRL, which introduces a novel reward decomposition into format, structure, and perceptual quality. This design explicitly enforces executable pipelines and logical reasoning order, which is a critical yet overlooked requirement for audio restoration agents.

\item \textbf{A audio perception reward design for audio restoration.} We introduce a novel reward design that combines modern objective audio metrics, serving as an effective surrogate for perceptual quality and promoting stable and efficient APRL optimization.

\item \textbf{AcoustBench:}  It comprises 88.9K instruction-response pairs with explicit chain-of-thought (CoT) annotations and executable tool chains, specifically curated to support training and evaluation of scheduling policies under coupled distortions.

\end{itemize}

\section{Related Work}

\textbf{Speech Enhancement.} Current research on audio enhancement is largely conducted within a single-task paradigm using task-specific datasets, where models are trained and evaluated separately for subtasks such as denoising \cite{lu2023mp,zhao2024mossformer2,wang2025mamba}, dereverberation \cite{lu2023mp,lay2024analysis}, speech-overlap suppression \cite{xu2024tiger}, source separation \cite{zhao2025mossformer2,xu2024tiger,zhao2025clearervoice}, audio restoration \cite{lemercier2024diffusion}, and target-speaker extraction \cite{ju2022tea,ju2023tea}. In real-world applications, enhancement requirements are often scenario-dependent and should be inferred from the input audio itself, thereby requiring adaptive combinations of multiple subtasks. Moreover, existing task-specific enhancement models still exhibit limited generalization, making it difficult to handle complex, coupled degradations caused by factors such as environmental variability and device mismatch. They also lack sufficient flexibility for practical audio shaping. To bridge this gap, the Deep Noise Suppression (DNS) \cite{reddy2020interspeech,ju2022tea,ju2023tea,dubey2024icassp} and URGENT challenges \cite{zhang2024neurips,zhang2025lessons,saijo2025interspeech,li2026icassp} have introduced the Personalized Speech Enhancement (PSE) track and an adaptive multi-distortion processing setting, respectively.





\textbf{LLM-Empowered Agent.} Recent studies\cite{patil2024gorilla,zhao2024diffagent,szot2025multimodal,yang2025embodiedbench}have highlighted the growth potential of multimodal large language models in terms of proficient tool use and decision-making in complex environments.MLLMs  have benefited from powerful reasoning capabilities and rich tools that interact with the environment\cite{hong2023metagpt,narayan2025introduction}. Embodiedbench: specifically evaluates the embodied task performance of multimodal large language models (MLLMs)\cite{yang2025embodiedbench}. Gorilla improved the LLM's responsiveness to tool calls by constructing datasets and fine-tuning\cite{patil2024gorilla}.In the vision domain, various methods have been proposed to enhance large models for vision tasks, such as the Hugging Face model \cite{shen2023hugginggpt}, image degradation\cite{lin2025jarvisir}, image art creation\cite{lin2025jarvisart,feng2026gen}, and the Azure model\cite{yang2023mm}.Audio agents have not yet been developed.

\textbf{Agentic Reinforcement Learning.} 
Reinforcement learning (RL) plays a crucial role in aligning (multimodal) large language models \cite{guo2025deepseek,xue2025dancegrpo,xu2025qwen3} with human preferences.
Reinforcement Learning from AI Feedback (RLAIF) \cite{gao2023scaling,team2023gemini,anthropic2024claude}has emerged as an important paradigm for LLM alignment. In audio evaluation tasks, AI-generated feedback can effectively approximate human judgments, significantly reducing reliance on manual annotation while maintaining evaluation quality, thereby substantially lowering the cost of dataset construction. Building on this, RLAIF learns reward functions from AI-generated feedback and optimizes LLMs using reinforcement learning methods such as proximal policy optimization (PPO)\cite{schulman2017proximal}and GRPO\cite{guo2025deepseek}. Unlike typical agents, our two-stage fine-tuning process integrates audio and language modalities, without involving visual modalities.

\section{Method}
We first present the overall workflow of StrixAE in Sec.~\ref{sec3.1}. We then describe the data generation pipeline used to construct AcoustBench, a high-quality audio dataset comprising cue samples and inference samples for audio enhancement tasks, in Sec.~\ref{sec3.2}. Finally, we detail the core components of StrixAE, including its two-stage training process and the Agent-Orchestrated Multi-model Enhancement Protocol, in Sec.~\ref{sec3.3}.



\subsection{Overview}
\label{sec3.1}
StrixAE is an audio enhancement model based on MLLM that can coordinate multiple audio enhancement models to process complex degraded audio data and achieve high-quality distortion audio enhancement. The StrixAE model operation mainly includes three steps: perceptual audio degradation inference, model selection and evaluation, and task execution and response generation, as shown in  Figure \ref{fig3}.
\begin{equation}
f(Q, \mathcal{A}) \rightarrow \mathcal{T} = \{ t_1, t_2, \dots, t_n \}
\label{eq:strixae}
\end{equation}

where $\mathcal{I}$ denotes the user  instruction , $\mathcal{A}$ represents the distorted audio, $\mathcal{T}$ represents the normalized tool call sequence, where each $t_i$ represents a specific audio enhancement operation. Finally, the enhanced output is obtained by:$\mathcal{A}_{\text{edit}} = G\left(\mathcal{A}, \mathcal{T}\right)$. $G(.)$  represents the audio enhancement environment.


\begin{figure}[t]
  \centering
  \centerline{\includegraphics[width=0.95\columnwidth]{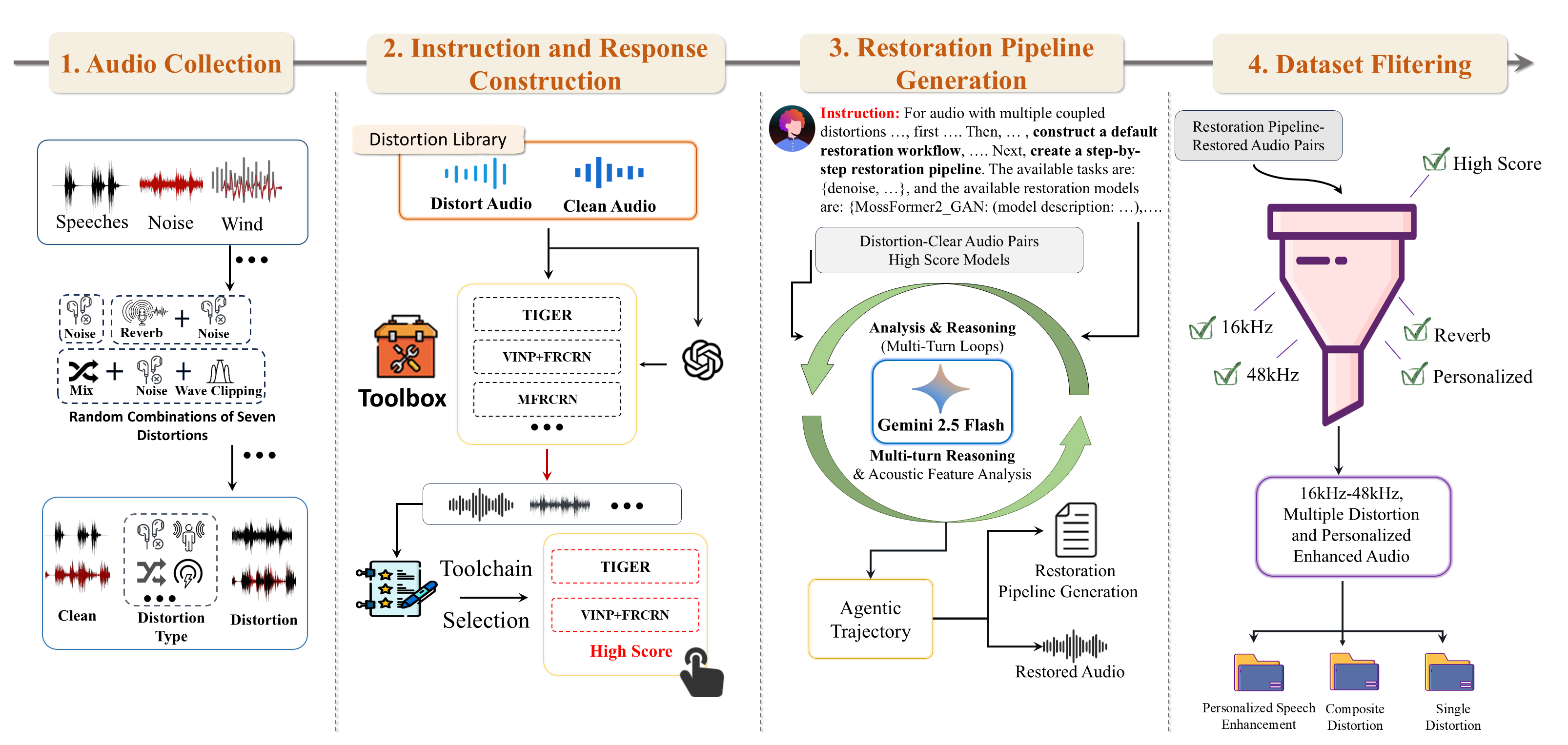}}
  \caption{An illustration of our data curation pipeline.}
  \label{fig2}
\end{figure}

\subsection{Dataset Construction}
\label{sec3.2}

We construct \textbf{AcoustBench}, a novel audio enhancement benchmark with explicit chain-of-thought (CoT) annotations, via a four-stage data generation pipeline (\ref{fig2}). Each dataset sample is defined as a 5-tuple:
$\langle \mathcal{A}, \mathcal{A}_{\text{tgt}}, \mathcal{I}, C, \mathcal{T} \rangle$,
where  $\mathcal{A}_{\text{tgt}}$ denotes its clean target audio, $C$ represents the step-by-step CoT reasoning.

\textbf{Audio collection.} We use real and clean audio datasets from the ICASSP 2023 DNS Challenge \cite{dubey2024icassp} and the 2025 URGENT Challenge \cite{saijo2025interspeech,li2026icassp}. These datasets include 350 hours of clean audio, 181 hours of real noisy audio \cite{thiemann2013diverse,fonseca2017freesound,gemmeke2017audio}, and reverberation data \cite{allen1979image}. Based on these resources, we construct 190 hours of paired audio with composite distortions. Further details are provided in Supplementary Material.

\textbf{Instruction and Response Construction.} Inspired by self-guided strategies~\cite{wang2023self}, we first designed 10 initial instructions as seeds. Based on these seeds, we used GPT-5.3 to generate 20 refined instruction candidates. These candidates were then manually reviewed and selected to remove ambiguous, duplicated, or low-quality samples, resulting in a curated set of high-quality instructions. Furthermore, for each audio sample in the training set, we designed two recommended tool invocation paths for audio restoration. These candidate paths were evaluated based on the quality of their processed outputs, using DNSMOS~\cite{reddy2022dnsmos} and ESTOI~\cite{jensen2016algorithm} as the assessment criteria. The selected paths were then paired with the corresponding clean and distorted audio samples to form the instruction-toolchain annotations. 



\textbf{Restoration Pipeline Generation.} Based on the generated instructions and responses, we obtain instruction--toolchain response pairs and corresponding audio pairs. Specifically, Gemini-2.5-Flash \cite{comanici2025gemini} is used to generate detailed logical chain responses, which typically include \textit{acoustic feature analysis}, \textit{problem diagnosis and reasoning}, \textit{restoration logic reasoning}, and a \textit{recommended restoration pipeline}. The restoration tasks and tool models are sourced from GitHub or Hugging Face.

\textbf{Dataset Filtering.}
To accommodate diverse real-world scenarios and enhance distorted audio at different sampling rates, we select high-quality instruction--response pairs of various types to satisfy scenario-specific requirements. AcoustBench contains 89K instruction--response pairs for the initial instruction-tuning stage. Further details are provided in Supplementary Material.

\begin{figure*}[t]
  \centering
  \includegraphics[width=0.9\textwidth]{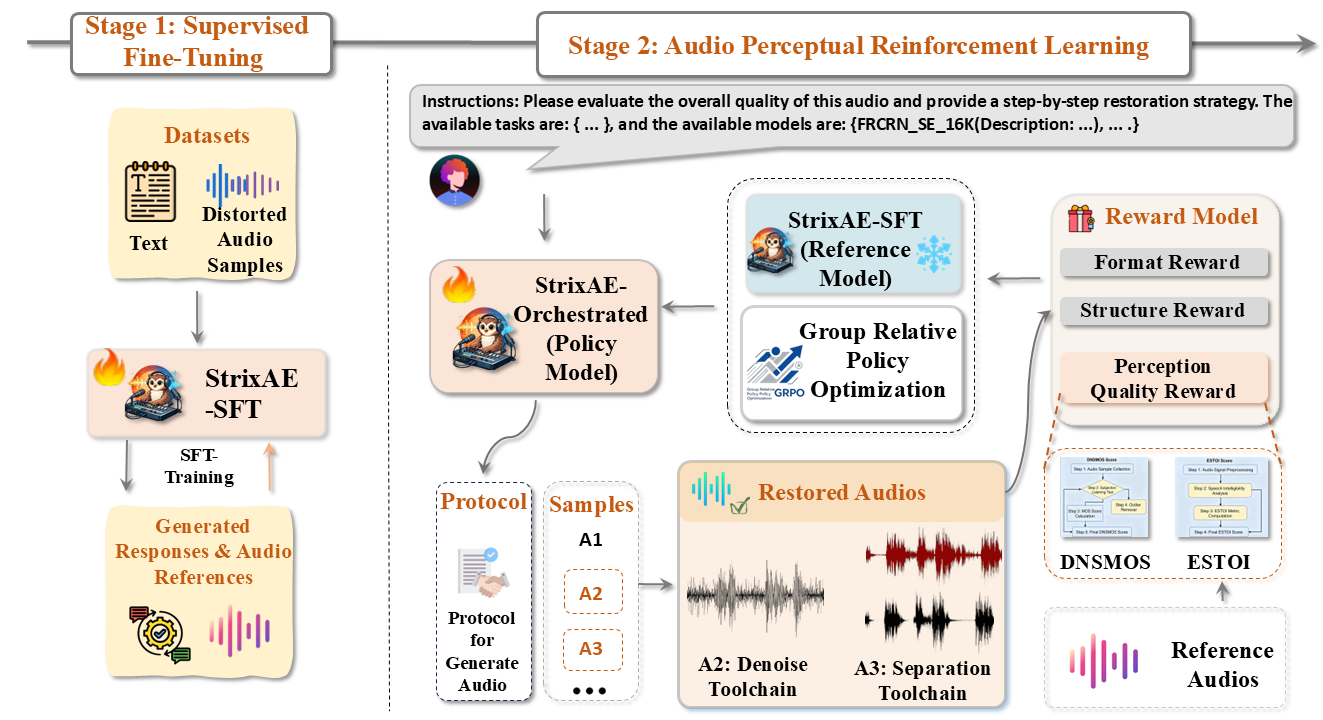}
  \caption{StrixAE employs a two-stage training framework. Initially, StrixAE undergoes \textbf{Supervised Fine-Tuning} on AcoustBench synthesized data, following user instructions and enhancing distorted audio. In the second stage, the  Audio Perceptual Reinforcement Learning \textbf{(APRL)} algorithm is applied to further enhance the robustness of StrixAE's enhancement system and improve its adaptability to audio enhancement in various real-world scenarios.}
  \label{fig3}
\end{figure*}


\subsection{StrixAE Framework}
\label{sec3.3}


\subsubsection{CoT Supervised Fine-tuning}
We use the AcoustBench dataset to perform SFT to obtain the StrixAE-SFT model. Our multimodal instruction samples can be represented as triples $($$\mathcal{I}$, $\mathcal{A}$, $\mathcal{R}$$)$. MLLM outputs the analysis process and the tool call chain based on the human instruction and the distorted audio: $\mathcal{R} = f \bigl(\mathcal{I}, \mathcal{A}; \theta\bigr)$.

\begin{equation}
L_{sft} = -\sum_{i=1}^N \log P_{\pi} \bigl( \mathcal{R}_i \mid \mathcal{I}_i, \mathcal{R}_i, \mathcal{R}_{<i}; \theta \bigr)
\label{eq:sft}
\end{equation}
Where $N$ represents the length of the actual response. The variables $\mathcal{I}$, $\mathcal{A}$, and $\mathcal{R}$ represent the user command, the distorted audio input, and the target response, respectively. Within this framework, $\mathcal{R}$ is composed of $C$ and $\mathcal{T}$.


\subsubsection{Audio Perceptual Reinforcement Learning}

Building on the SFT-initialized model, we further apply reinforcement learning to improve structural consistency, output validity, and perceptual awareness. Specifically, we design three interpretable and complementary rewards: a format reward $R_{\text{fmt}}$, which enforces a valid and executable pipeline structure; a structure reward $R_s$, which encourages the inclusion and correct ordering of key analytical sections; and a perceptual quality reward $R_q$, which evaluates the enhanced speech in terms of perceptual quality and intelligibility. The overall reward is defined as
\[
R = \lambda_f R_{\text{fmt}} + \lambda_s R_s + \lambda_q R_q,
\]
\noindent where $\lambda_f, \lambda_s, \lambda_q > 0$ are weighting hyperparameters that balance the contributions of the format, structure, and perceptual quality rewards.

\textbf{Format Reward.} The format reward is designed to ensure that the recommended restoration pipeline follows a valid and executable format. Let $y$ denote the tool output, $\mathcal{M}(y)$ the set of parsed tool names, and $\mathcal{V}$ the set of valid tools. The format reward is defined as:
\[
R_{\text{fmt}}(y)=
\begin{cases}
0, & \text{if the pipeline is missing or no valid steps can be parsed} \\
-0.25, & \text{if any parsed tool is invalid (}\mathcal{M}(y)\nsubseteq \mathcal{V}\text{)} \\
1, & \text{if all parsed tools are valid (}\mathcal{M}(y)\subseteq \mathcal{V}\text{)}
\end{cases}
\]

\textbf{Structure Reward.} 
We define a structure reward $R_s \in [0, 1]$ to encourage the model to generate responses with a complete and correctly ordered analytical structure.Let $y$ denote the generated response, which is expected to contain four target sections in a predefined order, denoted as $\{S_1, S_2, S_3, S_4\}$. We define indicator variables
\[
\delta_i(y) = 
\begin{cases}
1, & \text{if section } S_i \text{ appears in } y \text{ and follows the correct order}, \\
0, & \text{otherwise}.
\end{cases}
\]
The structure reward is computed as
\begin{equation}
R_s(y) = 
\begin{cases}
-1, & \text{if } S_4 \notin y, \\
\frac{1}{4} \sum_{i=1}^{4} \delta_i(y), & \text{otherwise}.
\end{cases}
\end{equation}

Sections that appear out of order do not contribute to the reward. The penalty term enforces the presence of the final section, ensuring completeness of the generated structure.

\textbf{Perception quality reward}. To evaluate the perceptual quality and intelligibility of enhanced speech, we design a composite reward based on two complementary objective metrics: DNSMOS and ESTOI. DNSMOS reflects perceptual speech quality, while ESTOI measures intelligibility. Let $d$ and $e$ denote the DNSMOS and ESTOI scores, respectively. We normalize each metric using a sigmoid function $\sigma(x)=1/(1+e^{-x})$. Specifically, we apply a sigmoid transformation with slope parameter $k_d$ and pivot $c_d$ to DNSMOS, and a sigmoid with slope $k_e$ and pivot $c_e$ to ESTOI:
\begin{equation}
s_d(d) = \frac{1}{1 + \exp\!\left(-k_d(d-c_d)\right)}, \qquad
s_e(e) = \frac{1}{1 + \exp\!\left(-k_e(e-c_e)\right)}.
\end{equation}

We then combine the two normalized scores using a weighted geometric mean with exponent $\alpha \in [0,1]$:
\begin{equation}
s_{\mathrm{joint}}(d,e)
= s_d(d)^{\alpha} \, s_e(e)^{1-\alpha}.
\end{equation}

Finally, the joint score is linearly rescaled to the interval $[-1,1]$:
\begin{equation}
R_q = 2 \, s_{\mathrm{joint}}(d,e) - 1.
\end{equation}

Substituting the above definitions yields:
\begin{equation}
R_q
=
2 \left[
\left(
\frac{1}{1 + \exp\!\left(-k_d(d-c_d)\right)}
\right)^{\alpha}
\left(
\frac{1}{1 + \exp\!\left(-k_e(e-c_e)\right)}
\right)^{1-\alpha}
\right]
-1 .
\label{eq:final_reward}
\end{equation}

\textbf{Novelty of our reward design.} While prior RL fine-tuning for LLMs often relies on a single task-specific reward (e.g., DNSMOS), audio restoration agents face unique challenges: the model must generate executable tool sequences and logically ordered analyses. Our reward function addresses this via three innovations: (1) a format reward that validates pipeline executability, (2) a structure reward that enforces section ordering, and (3) a joint perceptual reward that balances quality and intelligibility. To our knowledge, this is the first work to encode pipeline validity and reasoning order into RL rewards for audio enhancement agents.



\subsubsection{Agent-Orchestrated Multi-model Enhancement Protocol}
We build a service-oriented agent system on top of a modular pipeline framework for audio restoration. A lightweight HTTP interface enables remote execution and interaction during training. Concurrency is managed by a shared worker pool with a bounded queue controlled by semaphores, ensuring stable throughput and providing backpressure under high load. Tool usage is handled through a decoupled strategy that combines preloaded and resident tools, thereby reducing cold-start latency while limiting GPU memory consumption through a bounded cache. An executor module parses pipeline definitions and executes multi-step restoration workflows. Further details are provided in Supplementary Material.

\section{Experiment}
\subsection{Experimental Setup}
\textbf{Training Setup.} StrixAE is built upon Audio-Reasoner~\cite{xie2025audio} and fine-tuned with LoRA using the AdamW optimizer. In the SFT stage, the model is trained for 3 epochs with a batch size of 8 and a learning rate of $1 \times 10^{-5}$. During RL tuning, the sampling temperature is set to 1.0. We construct a joint reward model, defined in Equation~\ref{eq:final_reward}, using DNSMOS to assess perceptual quality and ESTOI to evaluate intelligibility. The overall reward is computed as a weighted combination with $\lambda_f = 1$, $\lambda_s = 0.3$, and $\lambda_q = 0.3$. Sigmoid normalization is applied with $(k_d, c_d) = (5, 2.5)$ for DNSMOS and $(k_e, c_e) = (8.0, 0.55)$ for ESTOI, and the two normalized scores are combined using $\alpha = 0.45$. All experiments are conducted on four NVIDIA H100 80G GPUs.

\textbf{AcoustBench Evaluation Dataset.} To comprehensively evaluate the performance of StrixAE, we develop the AcoustBench evaluation benchmark. This dataset integrates publicly available blind evaluation data from the 2020 DNS Challenge \cite{reddy2020interspeech}, the 2023 DNS Challenge \cite{dubey2024icassp}, and the 2025 URGENT Challenge \cite{saijo2025interspeech}. In addition, we introduce a new test dataset, \textbf{AcoustBench-Real}, collected from complex real-world environments. Further technical details are provided in Supplementary Material.

\textbf{Evaluation Metrics.} We use AcoustBench for training and tuning and evaluate our system on a real-world blind test set. To comprehensively assess audio quality, we employ multiple evaluation metrics, including DNSMOS \cite{reddy2022dnsmos}, NISQA \cite{mittag2021nisqa}, UTMOS \cite{saeki2022utmos}, and SCOREQ \cite{ragano2024scoreq}. Supplementary Material provides detailed explanations of these metrics and their corresponding evaluation tasks.



\textbf{Tool Settings.} 
This paper employs task-specific audio enhancement and audio separation models as external tools. We select lightweight and efficient models to ensure practical tool invocation; however, incorporating more advanced tools may further improve performance. Further details are provided in Supplementary Material.

\subsection{Experimental Results}

\definecolor{opensourcebg}{RGB}{250,248,228}
\definecolor{closesourcebg}{RGB}{235,247,235}

\begin{table*}[t]
\centering
\caption{Comparison of StrixAE with Single-Task and All-in-One methods on four real-world blind test datasets. $\uparrow$ denotes higher is better.}
\label{tab:comparison-1}
\small
\resizebox{\textwidth}{!}{
\setlength{\tabcolsep}{5pt}
\renewcommand{\arraystretch}{0.95}

\begin{tabular}{l cccc cccc}
\toprule

\multirow{2}{*}{\textbf{Method}}
& \multicolumn{4}{c}{\textbf{Real Recordings}}
& \multicolumn{4}{c}{\textbf{AcoustBench-Real}} \\
\cmidrule(lr){2-5}
\cmidrule(lr){6-9}
& DNSMOS $\uparrow$
& NISQA $\uparrow$
& UTMOS $\uparrow$
& SCOREQ $\uparrow$
& DNSMOS $\uparrow$
& NISQA $\uparrow$
& UTMOS $\uparrow$
& SCOREQ $\uparrow$ \\
\midrule

\rowcolor{opensourcebg}
\multicolumn{9}{c}{\textbf{\textit{Open-source Models}}} \\

TIGER \cite{xu2024tiger}
& 2.60 & 2.79 & 2.32 & 2.69
& 2.03 & 2.13 & 1.53 & 1.82 \\

MossFormer2\_SS\_16K \cite{zhao2025mossformer2}
& 2.78 & 2.60 & 2.23 & 2.56
& 2.10 & 1.83 & 1.53 & 1.74 \\

\midrule

MP-SENet \cite{lu2025explicit}
& 3.16 & 3.45 & 2.73 & 3.45
& 2.68 & 2.76 & 1.76 & 2.72 \\

MossFormer2\_GAN \cite{zhao2025clearervoice}
& 3.12 & 3.63 & 2.67 & 3.58
& 2.47 & 2.20 & 1.68 & 2.18 \\

TF-GridNet \cite{wang2023tf}
& 3.01 & 3.24 & 2.56 & 3.27
& 2.66 & 2.66 & 1.69 & 2.53 \\

\midrule

\textcolor{blue}{$\star$} StrixAE-SFT
& 3.12 & 3.61 & 2.58 & 3.26
& 2.72 & 2.78 & 1.84 & 2.49 \\

\textcolor{red}{$\star$} StrixAE-Orchestrated
& 3.18 & 3.67 & 2.84 & 3.59
& 2.71 & 2.79 & 1.83 & 2.54 \\

\midrule

\rowcolor{closesourcebg}
\multicolumn{9}{c}{
    \textbf{\textit{Close-source Models (Reference)}}
} \\

Audio Cleaner AI
& 2.94 & 3.47 & 2.59 & 3.25
& 2.33 & 2.61 & 1.63 & 2.19 \\

Adobe Acrobat
& 3.08 & 3.92 & 3.06 & 3.49
& 2.80 & 3.29 & 2.27 & 2.98 \\

\midrule
\addlinespace[0.3em]

\multirow{2}{*}{\textbf{Method}}
& \multicolumn{4}{c}{\textbf{Personalization Enhancement}}
& \multicolumn{4}{c}{\textbf{Blind Test Set}} \\
\cmidrule(lr){2-5}
\cmidrule(lr){6-9}
& DNSMOS $\uparrow$
& NISQA $\uparrow$
& UTMOS $\uparrow$
& SCOREQ $\uparrow$
& DNSMOS $\uparrow$
& NISQA $\uparrow$
& UTMOS $\uparrow$
& SCOREQ $\uparrow$ \\
\midrule

\rowcolor{opensourcebg}
\multicolumn{9}{c}{\textbf{\textit{Open-source Models}}} \\

TIGER \cite{xu2024tiger}
& 2.90 & 3.24 & 2.52 & 2.77
& 2.23 & 2.02 & 1.69 & 1.84 \\

MossFormer2\_SS\_16K \cite{zhao2025mossformer2}
& 2.98 & 3.17 & 2.65 & 2.93
& 2.33 & 1.62 & 1.58 & 1.67 \\

\midrule

MP-SENet \cite{lu2025explicit}
& 3.16 & 3.61 & 2.62 & 3.18
& 2.81 & 2.50 & 1.92 & 2.44 \\

MossFormer2\_GAN \cite{zhao2025clearervoice}
& 3.07 & 3.52 & 2.44 & 3.02
& 2.77 & 2.58 & 1.98 & 2.44 \\

TF-GridNet \cite{wang2023tf}
& 3.11 & 3.44 & 2.68 & 3.17
& 2.85 & 2.76 & 1.91 & 2.57 \\

\midrule

\textcolor{blue}{$\star$} StrixAE-SFT
& 3.23 & 3.98 & 2.78 & 3.23
& 2.77 & 2.71 & 1.79 & 2.15 \\

\textcolor{red}{$\star$} StrixAE-Orchestrated
& 3.24 & 3.98 & 3.06 & 3.49
& 2.83 & 2.74 & 1.98 & 2.36 \\

\midrule

\rowcolor{closesourcebg}
\multicolumn{9}{c}{
    \textbf{\textit{Close-source Models (Reference)}}
} \\

Audio Cleaner AI
& 3.06 & 3.47 & 2.68 & 3.14
& 2.60 & 2.53 & 1.89 & 2.35 \\

Adobe Acrobat
& 3.16 & 4.02 & 2.89 & 3.22
& 2.92 & 2.89 & 2.35 & 2.67 \\

\bottomrule
\end{tabular}
}
\end{table*}

\begin{table*}[t]
\centering
\caption{Evaluation results of top 3 systems on track 1 blind test set in URGENT Challenge 2025.}
\label{tab:urgent2025_top3_blind}
\footnotesize
\setlength{\tabcolsep}{2pt}
\begin{tabular}{l cc cccc cc}
\toprule
& \multicolumn{2}{c}{Non-intrusive}
& \multicolumn{4}{c}{Intrusive}
& \multicolumn{2}{c}{Downstream} \\
\cmidrule(lr){2-3} \cmidrule(lr){4-7} \cmidrule(lr){8-9}
Systems & DNSMOS $\uparrow$ & NISQA $\uparrow$
& PESQ $\uparrow$ & ESTOI $\uparrow$ & SDR $\uparrow$ & LSD $\downarrow$
& SpkSim $\uparrow$ & CAcc(\%) $\uparrow$ \\
\midrule
Rank 1 (Ours) & 2.88 & 3.22 & \textbf{2.64} & \textbf{0.82} & \textbf{12.66} & 2.93 & \textbf{0.76} & \textbf{79.80} \\
Rank 2        & 2.92 & 3.24 & 2.47 & 0.79 & 11.10 & 2.99 & 0.74 & 76.06 \\
Rank 3        & \textbf{2.94} & \textbf{3.25} & 2.45 & 0.79 & 11.25 & 3.66 & 0.71 & 77.09 \\
Rank 3        & \textbf{2.94} & \textbf{3.25} & 2.45 & 0.79 & 11.25 & 3.66 & 0.71 & 77.09 \\
Rank 3        & \textbf{2.94} & \textbf{3.25} & 2.45 & 0.79 & 11.25 & 3.66 & 0.71 & 77.09 \\
Rank 3        & \textbf{2.94} & \textbf{3.25} & 2.45 & 0.79 & 11.25 & 3.66 & 0.71 & 77.09 \\
Rank 3        & \textbf{2.94} & \textbf{3.25} & 2.45 & 0.79 & 11.25 & 3.66 & 0.71 & 77.09 \\

\midrule
Official Baseline & 2.85 & 2.77 & 2.24 & 0.76 & 10.24 & \textbf{2.72} & 0.70 & 75.60 \\
\bottomrule
\end{tabular}
\end{table*}

\subsubsection{Evaluation on AcoustBench Evaluation Dataset}
As shown in Table ~\ref{tab:comparison-1}, StrixAE achieves superior performance across four non-intrusive metrics. Compared to existing open-source methods, including both single-task and unified models. StrixAE outperforms most metrics and even surpasses several closed-source baselines (Adobe Acrobat and Audio Cleaner) in certain aspects.
Specifically, on the AcoustBench-Real dataset and its personalized enhancement test set, StrixAE achieves state-of-the-art (SOTA) results on DNSMOS, UTMOS, and SCOREQ. Against the strong baseline TF-GridNet\cite{wang2023tf}, StrixAE obtains significant improvements of 0.17, 0.43, 0.28 and 0.32, respectively. 

When handling real-world degradations, DNSMOS score of StrixAE is only 0.09 lower than that of the closed-source baseline. On the Blind Test challenge set, the gap with the leading baseline is further narrowed, with differences of only 0.02, 0.02, and 0.21 for DNSMOS, NISQA and SCOREQ, respectively.These cross-dataset comparisons demonstrate that StrixAE not only maintains high-fidelity enhancement but also exhibits strong generalization robustness against complex distortions.

\begin{figure*}[th]
  \centering
  \includegraphics[width=0.85 \textwidth]{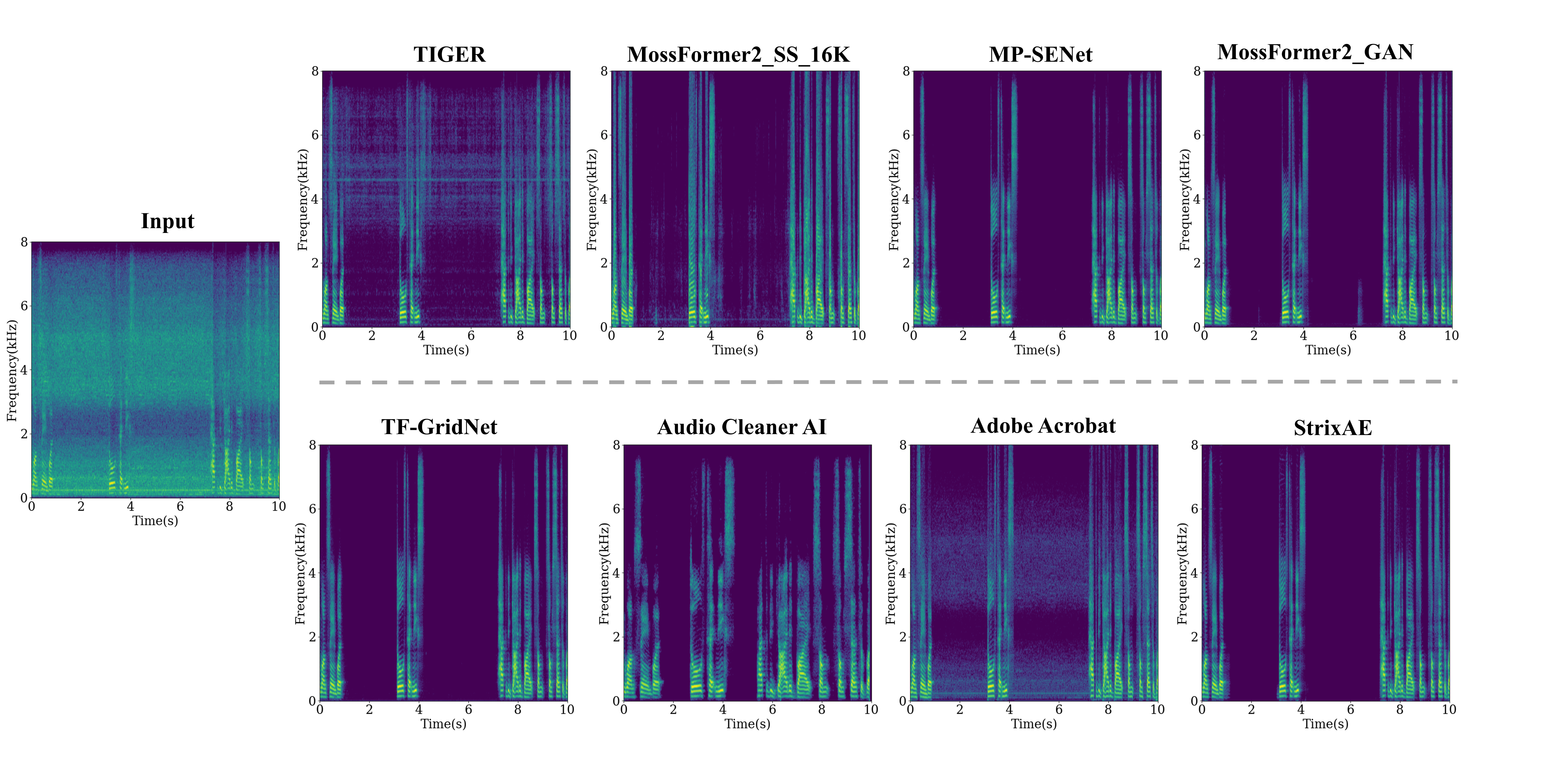}
  \caption{Spectrum visualization on the Real-Recording test dataset.}
  \label{fig4-visual}
\end{figure*}

As illustrated in Figure \ref{fig4-visual}, StrixAE effectively recovers speech frequencies compared to single-task methods. In contrast to other integrated models, it achieves better denoising performance while better preserving harmonic structures, thereby delivering high-quality audio enhancement. Additional results are provided in Supplementary Material.


Furthermore, the StrixAE-Orchestrated variant outperformed its SFT-based counterpart in most evaluation scenarios, validating the effectiveness of the proposed APRL fine-tuning strategy, which significantly improves generalization ability and decision accuracy in complex scenarios.

\subsubsection{Visualization of Reward Trends for APRL}
\begin{figure*}[th]
  \centering
  \includegraphics[width=0.95\textwidth]{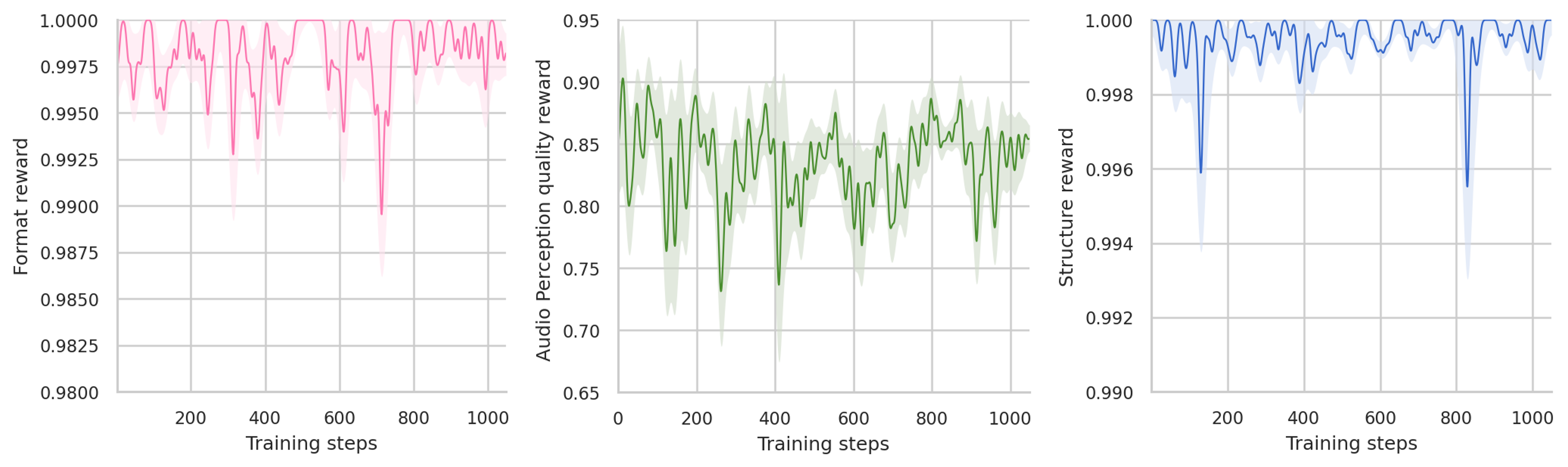}
  \caption{ Visualization of the reward trends across training steps of for StrixAE.}
  \label{fig4}
\end{figure*}

Figure~\ref{fig4} illustrates the training dynamics of APRL across three reward components. The format reward rapidly converges to near-optimal values during the early training stage and fluctuates only slightly thereafter, indicating that the model quickly internalizes the required output format. In contrast, the audio perception quality (APQ) reward exhibits greater variability and improves more gradually, reflecting a more complex optimization landscape that requires sustained exploration. Meanwhile, the structure reward rises quickly and remains consistently high, suggesting that the model benefits from structural priors acquired during SFT, analogous to inherited ``parameter preferences''.

This pattern indicates a sequential stabilization process: the model first optimizes easily regularized rewards, such as format and structure, before shifting capacity toward the more challenging APQ objective, whose search space is broader and less deterministic. Unlike reasoning-centric models such as DeepSeek-R1 \cite{guo2025deepseek}, no clear ``aha moment'' emerges during training.

Nevertheless, the training dynamics reveal a key strength of our design: the format and structure rewards stabilize rapidly, indicating that the model quickly internalizes the required output template — a prerequisite for reliable agent deployment. The slower improvement of the perceptual reward reflects the inherent difficulty of audio quality optimization, which our APRL continues to refine over time. This separation of concerns (format → structure → quality) is a deliberate innovation that makes training more stable than end-to-end perceptual-only RL.

\section{Ablation Study}
\textbf{Training strategy.} As shown in Table ~\ref{tab:comparison-1}, we conducted a three-task comparison of StrixAE with several baseline strategies to analyze the impact on task planning and model selection, including:
(I) the task type is accurately identified, while both task execution order and model selection are random;
(II) the task execution order is random, and the model is selected by StrixAE;
(III) the model is randomly selected, and the task execution order is predicted by StrixAE. 

\begin{table}[t]
\centering
\caption{Comparison of StrixAE with other strategies on the Real Recordings validation set.}
\label{tab:comparison-1}
\setlength{\tabcolsep}{5pt}
\begin{tabular}{lcccc}
\toprule
Strategy                          & DNSMOS $\uparrow$ & NISQA $\uparrow$ & UTMOS $\uparrow$ & ScoreQ $\uparrow$ \\
\midrule
(I) Random Order and Model        & 2.72 & 3.02 & 1.47 & 2.47 \\
(II) Random Order + Predict Model & 2.84 & 3.11 & 2.52 & 3.02 \\
(III) Random Model + Predict Order& 2.80 & 3.16 & 2.41 & 2.80 \\
\cellcolor[HTML]{EDF4FB}{\color{blue}$\star$} \textbf{StrixAE-SFT} 
& \cellcolor[HTML]{EDF4FB}\textbf{3.12} 
& \cellcolor[HTML]{EDF4FB}\textbf{3.61} 
& \cellcolor[HTML]{EDF4FB}\textbf{2.58} 
& \cellcolor[HTML]{EDF4FB}\textbf{3.26} \\
\cellcolor[HTML]{DCEAF6}{\color{red}$\star$} \textbf{StrixAE-Orchestrated} 
& \cellcolor[HTML]{DCEAF6}\textbf{3.18} 
& \cellcolor[HTML]{DCEAF6}\textbf{3.67} 
& \cellcolor[HTML]{DCEAF6}\textbf{2.84} 
& \cellcolor[HTML]{DCEAF6}\textbf{3.59} \\
\bottomrule
\end{tabular}
\end{table}

\textbf{Results.} Table \ref{tab:comparison-1} quantifies the individual and combined effects of task scheduling and model selection. Both learned model selection (II) and task order prediction (III) outperform the fully random setting (I), verifying the necessity of each module. Integrating these two designs further boosts performance in StrixAE-SFT. By jointly optimizing task planning and dynamic model routing, StrixAE-Orchestrated attains the optimal results across all speech quality metrics.

Table~\ref{tab:comparison-2} compares different training strategies. We observe that using only SFT already yields competitive performance across all metrics. However, applying reinforcement learning (RL) alone results in a noticeable performance degradation, indicating that RL by itself is insufficient for stable optimization. By combining SFT with RL, our method achieves the best performance across all evaluation metrics, demonstrating the effectiveness of the proposed two-stage training strategy.

\begin{table}[th]
\centering
\caption{Evaluation results on the urgent real-world blind test dataset compared with other strategies.}
\label{tab:comparison-2}
\setlength{\tabcolsep}{5pt}
\begin{tabular}{lcccc}
\toprule
Strategy & DNSMOS $\uparrow$ & NISQA $\uparrow$ & UTMOS $\uparrow$ & ScoreQ $\uparrow$ \\
\midrule
(I) Only SFT          & 3.12 & 3.61 & 2.58 & 3.26 \\
(II) Only RL          & 3.11 & 3.54 & 2.53 & 3.22 \\
\cellcolor[HTML]{E6E6E6}(III) SFT + RL  
& \cellcolor[HTML]{E6E6E6}\textbf{3.18}
& \cellcolor[HTML]{E6E6E6}\textbf{3.67}
& \cellcolor[HTML]{E6E6E6}\textbf{3.84}
& \cellcolor[HTML]{E6E6E6}\textbf{3.59} \\
\bottomrule
\end{tabular}
\end{table}


\section{Conclusion} In this paper, we propose a novel two-stage training framework for building audio enhancement agents, and introduce StrixAE, an agent trained under this framework. Our main contributions include: (1) APRL, a reinforcement learning fine-tuning method that employs structured rewards to explicitly enforce pipeline executability and logical segmentation order; (2) a decomposed reward design that separately considers format validity, inference structure, and perceptual quality, improving training stability compared to single perceptual rewards; and (3) AcoustBench, the first large-scale benchmark specifically designed for learning and evaluating audio agent scheduling strategies under complex distortion coupling.  Extensive experiments demonstrate that StrixAE outperforms most open-source and proprietary solutions, exhibiting superior generalization robustness. Across various real-world datasets and personalized enhancement tasks, it achieves state-of-the-art performance on multiple perceptual metrics.

\newpage
\bibliography{reference}
\bibliographystyle{abbrvnat}

\newpage

\clearpage

\end{document}